\documentclass[aps,prl,twocolumn,amsmath,
showpacs]{revtex4}
\usepackage{epsfig}

\begin{document}
\title{Collective excitations of trapped Fermi or Bose gases}
\author{Andr\'as Csord\'as}
\email{csordas@tristan.elte.hu}
\affiliation{Research Group for Statistical Physics of the
             Hungarian Academy of Sciences,
             P\'azm\'any P. S\'et\'any 1/A, H-1117 Budapest, Hungary}
\author{Zolt\'an Adam}
\email{adamzoltan@complex.elte.hu}
\affiliation{Department of Physics of Complex Systems, E\"otv\"os University,
             P\'azm\'any P. S\'et\'any 1/A, H-1117 Budapest, Hungary}
\date{\today}
\begin{abstract}
A new method is developed to calculate all excitations 
of trapped gases using hydrodynamics
at zero temperature for any equation of state $\mu=\mu(n)$ and for 
any trapping potential.
It is shown that a natural scalar product can be defined for the mode
functions, by which the wave operator is hermitian and the mode 
functions are orthogonal.
It is also shown that the
Kohn-modes are exact for harmonic trapping in hydrodynamic theory.
Excitations for fermions are calculated in the 
BCS-BEC transition region using the equation of
state of the mean-field Leggett-model for isotrop harmonic trap potential.  
\end{abstract}
\maketitle

Several experiments on trapped ultracold gases  
probed in the past decade the collective excitations of atomic gases. 
Earlier measurements
on bosons \cite{Jin97,Mewes96} and more recent measurements on fermions
\cite{kinast04,bartenstein042} near Feshbach resonances can be 
explained rather satisfactorily
using hydrodynamics at zero temperature. In his seminal
paper \cite{Stringari96} Stringari applied first hydrodynamics for trapped
bosons undergoing Bose-Einstein condensation. His predictions were 
confirmed by experiments \cite{Jin97,Mewes96}. Later,
using the same approach he predicted \cite{Stringari04} also the qualitative
behavior of low lying modes 
for fermions in the whole crossover region from a BCS type superfluid
Fermi gas to a molecular Bose-Einstein condensation (BEC)
\cite{leggett80,Engelbrecht97,Marini98}. 
Now, several recent theoretical papers appeared in the
literature \cite{Fuchs03,Heiselberg04,Hu04,Bulgac05,Kim04,Astrakharchik05} 
using hydrodynamic theory to better explain
the measurements on the BCS-BEC transition. 
In general, no exact
solution to the hydrodynamic equations are known, except when
the equation of state has the polytropic form $\mu(n)\propto n^\gamma$
\cite{Heiselberg04}.

The hydrodynamic approach leads to a wave 
equation for the density oscillations.
In principle, this wave equation can be solved for a single oscillating
mode, if the boundary conditions for the density oscillations are known.
Bulgac et al. \cite{Bulgac05} has written the eigenvalue equation for
a single mode in such a way that the two sides were hermitian,
but did not address the question of the function space to which
all the excitations should belong.
Here we use a different approach.
For a general equation of state $\mu=\mu(n)$
it is usually very difficult to prescribe 
appropriate boundary conditions at the surface of the gas.
There are a few examples where this problem is circumvented
using some ansatz on the spatial forms of the excitations 
\cite{Hu04,Astrakharchik05}. Here we shall introduce a natural
scalar product, by which
the wave operator itself is hermitian and
the boundary conditions can be treated as in 
quantum-mechanical problems: the mode functions are
square integrable functions. The scalar product we shall use automatically
ensures particle conservation. By this way  finding excitation frequencies
are relatively easy: the task is to calculate matrix-elements
of the wave operator with the natural scalar product, and then calculate
the eigenvalues of the resulting matrix. We shall demonstrate the whole
procedure for the mean-field model of the BCS-BEC transition for isotropic
trap potential and compare our results with that of the scaling ansatz 
approach for the lowest $l=0$ monopole mode.

In hydrodynamic theory for  trapped gases at zero temperature
density oscillations are given by the continuity equation
\begin{equation}
\frac{\partial n}{\partial t} +\nabla(n {\bf u})=0,
\label{eq:cont_eq}
\end{equation}
and the Euler-equation
\begin{equation}
\frac{\partial {\bf u}}{\partial t}+{\bf u}\nabla{\bf u}=
-\frac{1}{mn}(\nabla P)-\frac{(\nabla V)}{m},
\label{eq:euler}
\end{equation}
where $\bf u$ is the velocity field, $n$ is the density, $t$ is the time,
$P$ is the pressure, $m$ is the particle mass, and $V$ is the 
external trapping potential.

Knowing the
equation of state $\mu=\mu(n)$ at zero temperature in the corresponding
homogeneous system
the equilibrium density in the trapped case can be
determined from the local chemical potential
\begin{equation}
\mu=\mu({\bf r})\equiv\mu(n_0({\bf r}))=\mu_0-V({\bf r}),
\label{eq:loc_chem_pot}
\end{equation}
where $\mu_0$ is the overall constant chemical potential. 
For confining potentials the solution of this equation for
positive $n_0(\mathbf{r})$ supplies an equilibrium density, 
which has a finite support with a well defined boundary. Typically
$n_0$ decreases to zero by approaching the boundary.
Using the thermodynamic identity 
\begin{equation}
\frac{\partial \mu}{\partial n} =\frac{1}{n}\left(
\frac{\partial P}{\partial n} \right)
\end{equation}
valid also at $T=0$ the gradient of Eq. (\ref{eq:loc_chem_pot})
gives the important vector identity
\begin{equation}
-A_0({\bf r})\frac{\nabla n_0({\bf r})}{n_0({\bf r})}=\nabla V({\bf r}),
\quad 
A_0({\bf r})\equiv \left(\frac{\partial P}{\partial n} \right)_{n=n_0(\bf r)}
\label{eq:imp_id}
\end{equation}
Eq. (\ref{eq:imp_id}) plays a central role in the following in simplifying the
linearized hydrodynamics.
In mechanical equilibrium the equilibrium pressure $P_0$ must satisfy
\begin{equation}
\nabla P_0({\bf r})=-n_0({\bf r})\nabla V({\bf r}),
\label{eq:Archimedes}
\end{equation}
otherwise the right hand side of Eq. (\ref{eq:euler}) will not vanish
for $\mathbf{u}=0$ (this is the local form of the Archimedes-law for
a general external potential).
Close to equilibrium $\bf u$ and 
$\delta n({\bf r},t)=n({\bf r},t)-n_0({\bf r})$
are small,   and $P$ can be expanded 
to first order in $\delta n$ as 
\begin{equation}
P({\bf r},t)=P_0({\bf r})+
\left(\frac{\partial P}{\partial n}\right)_0\delta n({\bf r},t).
\end{equation}
An important restriction for $\delta n$ is particle conservation 
$\int d^3r\, \delta n(\mathbf{r},t)=0$.
Linearizing the continuity equation (\ref{eq:cont_eq}) and the Euler-equation
(\ref{eq:euler}) in $\delta n$ and $\bf u$ using the local form of
the Archimedes-law (\ref{eq:Archimedes}) and the identity (\ref{eq:imp_id})
the linearized hydrodynamic equations can be written as
\begin{equation}
\frac{\partial \delta n}{\partial t} +\nabla(n_0 {\bf u})=0,
\label{eq:cont_eq_lin}
\end{equation}
\begin{equation}
\frac{\partial {\bf u}}{\partial t}=
-\nabla\left[\frac{A_0}{n_0m} \delta n\right].
\label{eq:euler_lin}
\end{equation}

Let us  introduce a new field by
\begin{equation}
\Psi({\bf r},t)=\sqrt\frac{A_0({\bf r})}{n_0({\bf r})}\delta n({\bf r},t)
\label{eq:Psi_def}
\end{equation}
where $\Psi$ has the same support as $n_0$, $A_0$ and $\delta n$.
From now on, we allow complex fields $\Psi$ 
(which are more convenient for problems,
where angular momentum is conserved).
Eliminating $\mathbf{u}$ from Eqs. (\ref{eq:cont_eq_lin}) and (\ref{eq:euler_lin})
a wave equation 
\begin{equation}
\frac{\partial^2 \Psi}{\partial t^2}+{\hat G}_\Psi \Psi=0,
\label{eq:waveeqforpsi}
\end{equation}
can be derived for $\Psi$,
where ${\hat G}_\Psi$ is given by
\begin{equation}
\hat{G}_\Psi=-\sqrt\frac{A_0({\bf r})}{n_0({\bf r})}\cdot
\nabla \cdot \frac{n_0({\bf r})}{m} \cdot \nabla \cdot 
\sqrt\frac{A_0({\bf r})}{n_0({\bf r})}.
\label{eq:wave_op}
\end{equation}
The main advantage of the field $\Psi$ 
is that its wave operator $\hat{G}_\Psi$
is manifestly
hermitian \cite{hermit} with respect to the scalar product
\begin{equation}
\langle \Psi_1 | \Psi_2 \rangle=\int_{n_0({\bf r})>0} d^3r\, \Psi_1^*({\bf r})
\Psi_2({\bf r})
\label{eq:scalproddef}
\end{equation}
The scalar product (\ref{eq:scalproddef}) is trivially the correct one for a 
homogeneous system with periodic boundary conditions. It was used 
for a weakly interacting trapped Bose-gas \cite{Csordas99}, where 
$\mu(n)\propto n$.
The same idea of finding a proper scalar product
for eigenmodes of a trapped, noninteracting Bose or Fermi gas
at finite temperature using the hydrodynamic approach was applied
in \cite{Csordas01}. A single eigenmode 
\begin{equation}
\Psi_i({\bf r},t)=\sin (\omega_i t+\phi_0)\Psi_i({\bf r}),
\end{equation}
fulfills the eigenvalue equation
\begin{equation}
\omega^2_i\Psi_i({\bf r})=\hat{G}_\Psi \Psi_i({\bf r}).
\label{eq:eigenvaleq}
\end{equation}
Solutions of (\ref{eq:eigenvaleq}) are (or for degenerate
eigenvalues can be) orthonormalized with the scalar product 
(\ref{eq:scalproddef}) 
\begin{equation}
\delta_{ij}=\int_{n_0({\bf r})>0} d^3r\, \Psi_i^*({\bf r})
\Psi_j({\bf r}).
\label{eq:orhtogonality}
\end{equation}
$\Psi_0(\mathbf{r})=\textrm{Const}\cdot\sqrt{n_0(\mathbf{r})/A_0(\mathbf{r})}$ 
is always
a solution to (\ref{eq:eigenvaleq}) with $\omega_0=0$. Eq. 
(\ref{eq:Psi_def}) implies that $\delta n_i=\Psi_0 \Psi_i$, thus
the orthogonality relation (\ref{eq:orhtogonality}) shows
that all the modes with $i\ne 0$ are automatically particle conserving,
and the mode $\Psi_0$ should be cancelled from the solutions.

Taking a complete orthonormal basis, i.e.,
$\delta_{i,j}=\langle \varphi_i |\varphi_j \rangle$
the squared excitation frequencies $\omega^2$ can be obtained from the
eigenvalues of the matrix
\begin{equation}
G_{i,j}=\langle \varphi_i |\hat{G}_\Psi |\varphi_j \rangle.
\label{eq:gmatrixdef}
\end{equation}
The matrix elements in (\ref{eq:gmatrixdef}) 
require the knowledge or the
numerical evaluation of spatial derivatives of the basis functions.
Usually this causes big numerical errors because the high lying
modes are rapidly oscillating functions.
In practice, it is much
better to apply the spatial derivatives to the (spatially varying) 
coeffectients of the wave equation, which are usually not oscillating
too much.

The wave operator (\ref{eq:wave_op}) has the structure 
$\hat{G}_\Psi=-R \nabla Q\nabla R$. If there exists a similar system
for which the boundary is the same and
the wave operator has also the structure 
$\hat{G}^0=-R_0 \nabla Q_0\nabla R_0$ but with known spectra and
eigenfunctions
\begin{equation}
\hat{G}^0 \varphi_i=\epsilon_i^{(0)}\varphi_i
\end{equation}
then one can eliminate the unwanted spatial derivatives of the basis functions
in the matrix elements if the basis is given by $\varphi_i$, $(i=0,1,\ldots)$.
Let us introduce $\alpha$ and $\beta$ by
\begin{equation}
\alpha=\alpha(\mathbf{r})=Q/Q_0,\quad \beta=\beta(\mathbf{r})=R/R_0,
\end{equation}
then the matrix elements can be written as
\begin{equation}
G_{i,j}=\int d^3r \varphi_i^*(\mathbf{r}) \varphi_j(\mathbf{r})
G_{i,j}(\mathbf{r}),
\end{equation}
where
\begin{gather}
G_{i,j}(\mathbf{r})=\frac{\epsilon_i^{(0)}+\epsilon_j^{(0)}}{2}\alpha\beta^2+
R_0^2 Q_0\alpha(\nabla \beta)(\nabla \beta)\nonumber\\
+\frac{1}{2}R_0^2\nabla\left[Q_0 \beta^2(\nabla\alpha) \right].
\end{gather}

For an isotrop
harmonic trapping potential $V(\mathbf{r})=m\omega_0^2 r^2/2$ 
and for any equation of state $\mu=\mu(n)$ a 
whole series of exact solutions of the wave equation can be given.
If $\Psi(\mathbf{r})$ is chosen to be
\begin{equation}
\Psi({\bf r})=\mbox{Const}
\sqrt{\frac{n_0({\bf r})}{A_0({\bf r})}}
r^l Y_l^m  (\vartheta,\phi), \quad l>0
\end{equation}
then this mode function fulfills the wave equation with eigenvalue
$\omega^2=\omega_0^2l$. The three $l=1$ modes are the Kohn-modes
(see Ref. \cite{Bulgac05})
for isotropic trapping.

As  a specific, nontrivial model let us consider the
mean-field model of Leggett \cite{leggett80}
for the BCS-BEC transition. 
The Leggett  model  
is fixed in homogeneous systems  by the gap equation
\begin{equation}
\sum_{\bf k}\frac{1}{2}\left(
\frac{1}{E_{\bf k}}-\frac{1}{\varepsilon_{\bf k}}\right)=
-\frac{m}{4\pi\hbar^2a},
\label{eq:leg_gap_hom}
\end{equation}
and by the number equation
\begin{equation}
N=\sum_{\bf k}\left( 1-\frac{\varepsilon_{\bf k}-\mu}{E_{\bf k}} \right),
\label{eq:leg_n_hom}
\end{equation}
where $E_{\bf k}=\sqrt{(\varepsilon_{\bf k}-\mu)^2+\Delta^2}$,
$\varepsilon_{\bf k}=\hbar^2 k^2/(2m)$, 
$\Delta$ is the pairing gap and $a$ is the $s$-wave scattering length.
The equation of state $\mu=\mu(n)$ is implicitly given by the model. 
This model captures the essential features of the BCS-BEC transition.
However, recent Monte-Carlo data on the equation of state
show \cite{Astrakharchik05} that there are corrections 
to the mean-field results of the
Leggett model which should be taken into account for the equation of state,
especially close to unitarity (i.e., around the $a=\infty$ point).
Here we study the above model for simplicity.
In the trapped case we use the $\mu(n)$ function taken from the model
and solve (\ref{eq:loc_chem_pot}) for the density profile keeping 
$N=\int d^3r\, n(\mathbf{r})$ to be fixed. The equilibrium pressure for any
trap potential $V(\mathbf{r})$ can be calculated from the
local form of the Archimedes law (\ref{eq:Archimedes}). Knowing
the pressure and the density the calculation of $A_0(\mathbf{r})$ 
with help of (\ref{eq:imp_id}) is straightforward.
The details of the full calculation for the Leggett model will be published 
elsewhere \cite{Adamunp}.

\begin{figure}
\centerline{
\epsfig{file=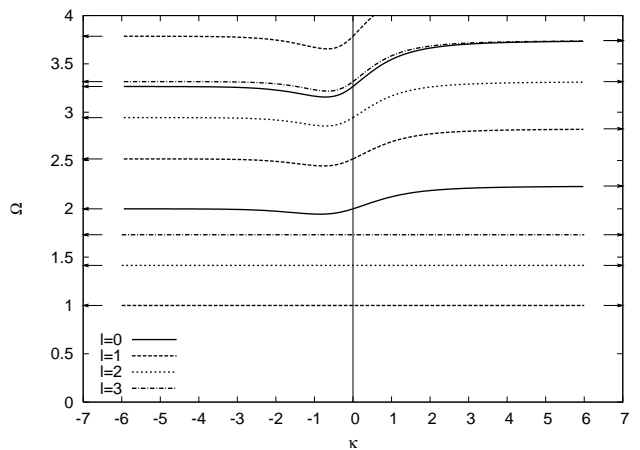,height=\hsize,angle=270}
}
\caption{Excitation frequencies 
$\Omega=\omega/\omega_0$ of the Leggett model as a function
of $\kappa=d/(aN^{1/6})$ on both sides of the BCS-BEC transition for isotrop
harmonic trapping. Arrows on the left (right) denote the excitations
for a non-interacting Fermi gas (for a weakly interacting Bose gas). 
\label{fig:whole_sp}}
\end{figure}

For isotrop harmonic trapping there is a dimensionless coupling
parameter: $\kappa=d/(aN^{1/6})$, where $d=\sqrt{\hbar/m\omega_0}$
is the oscillator length. The spectra
depends only on $\kappa$. In three cases the spectra is exactly
known \cite{Fuchs03,Heiselberg04,Stringari04} 
because the equation of state  has a polytropic form:
$\mu\propto n^\gamma$. These particular values are $\kappa=-\infty$
(BCS limit), $\kappa=0$ (unitarity limit) and $\kappa=\infty$ 
(BEC limit). In these cases all the mode functions can be constructed
exactly \cite{Fuchs03}, even in the nonisotropic case (The methods
of Refs. \cite{Csordas99,Csordas00} can be easily 
employed to the  polytropic equation of state).
We used the $\kappa=-\infty$ mode functions \cite{Bruun99} on the
BCS-side and the $\kappa=\infty$ mode functions \cite{Heiselberg04}
on the BEC-side as basis functions. 
Our numerical results for different angular momentum $l$ can
be seen on Fig. \ref{fig:whole_sp}.
Arrows on both sides show the limiting well-known collective
oscillation frequencies \cite{Stringari04}. 
\begin{figure}
\centerline{
\epsfig{file=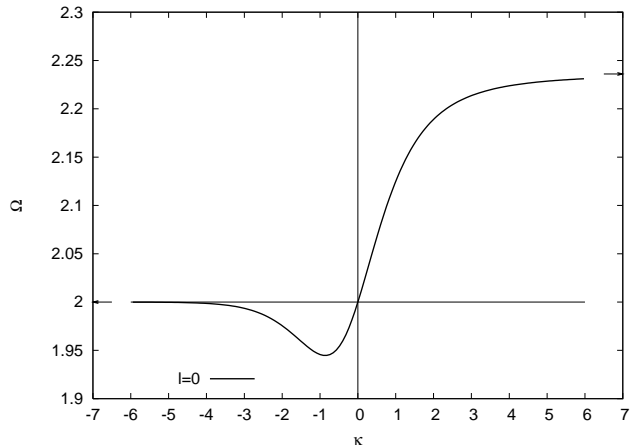,height=\hsize,angle=270}
}
\caption{The lowest $l=0$ excitation frequency $\Omega=\omega/\omega_0$
as a function
of $\kappa=d/(aN^{1/6})$ on both sides of the BCS-BEC transition for isotrop
harmonic trapping. Arrow on the left (right) denotes the excitation
for a non-interacting Fermi gas (for a weakly interacting Bose gas). 
\label{fig:kvadrupolmode}}
\end{figure}
In Fig. \ref{fig:kvadrupolmode} the behavior of the
lowest $l=0$ quadrupole mode can be seen as a function of $\kappa$. This mode
is the lowest $\kappa$ dependent mode on  Fig. \ref{fig:whole_sp}. 
The scaling ansatz
approach \cite{Hu04,Astrakharchik05} gives quite a good result
for this particular mode. On the scale of Fig. \ref{fig:kvadrupolmode}
the two curves would be practically indistinguishable. 

\begin{figure}
\centerline{
\epsfig{file=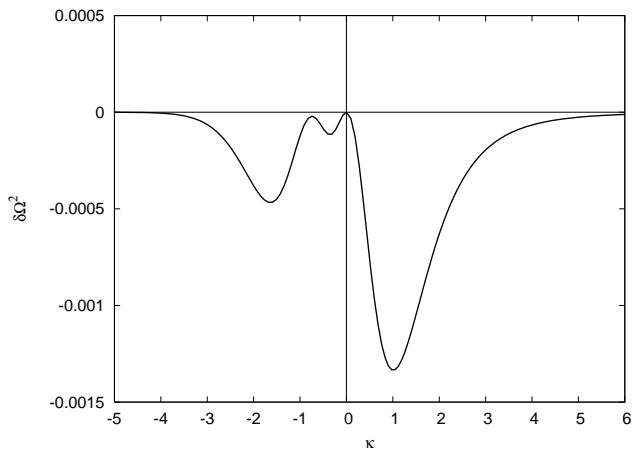,height=\hsize,angle=270}
}
\caption{Comparison of the exact frequencies and those of given by
the scaling ansatz as a function of $\kappa=d/(aN^{1/6})$ for the lowest
$l=0$ monopol mode.
$\delta \Omega^2$ defined as 
$\delta \Omega^2=(\omega^2-\omega^2_{sc})/\omega^2_0$ 
(See text for  $\omega^2_{sc}$).
\label{fig:scaling}
}
\end{figure}

In order to give a quantitative measure about the quality of the
latter approach
we compare in Fig. \ref{fig:scaling} our excitation frequencies with
those given by the scaling ansatz.
In the isotrop harmonic trapping potential case $\omega_{sc}^2$ is given
\cite{Hu04,Astrakharchik05} by $\omega_{sc}^2/\omega^2_0=9
\langle n\partial \mu/\partial n \rangle/(2\langle V\rangle)-1,$
where the average of a quantity like $V$  is taken as 
$\langle V \rangle = \int d^3r\, V(\mathbf{r})n_0(\mathbf{r})$.
From the figure it is clearly seen that scaling ansatz is exact
at $\kappa=0,\pm \infty$, but between these values it is not.
However, the difference in the isotropic case 
is so small that is much less than the
experimental resolution.
We have preliminary data \cite{Adamunp} for the excitation
frequencies for the experimentally relevant
axially symmetric harmonic trapping as well.
Once again the scaling ansatz differs a little for the radial and axial
quadrupole modes for a general intermediate coupling $\kappa$.
For such large anisotropies as in \cite{kinast04,bartenstein042} 
however, the deviation $\delta \Omega^2$ is much bigger
than in the isotropic case.

Finally, let us turn to the conclusions. We gave a straightforward method
how to solve the hydrodynamic equations in the trapped case
if the equation of state is known. We introduced a natural scalar
product for the transformed wave operator by which the operator is
hermitian. Collective excitations by our method can be found by
simply diagonalizing the matrix of the wave operator on some basis.
The power of our method lies in the fact that we calculate
the whole spectra. We can predict the behavior
of the excitations also for those modes for which no scaling
ansatz is known. The method is not limited to a particular
trap potential (isotropic or not), nor a given mode. 
There is no additional approximation, the method calculates
the (numerically) exact modes given by the hydrodynamic
theory. 
Here we wanted to present the basic formalism and applied to the simplest
isotrop harmonic case. We compared the numerically 
exact excitation frequencies with that of the
scaling ansatz for the monopole mode and showed that the difference
between the two frequencies are extremely small in the whole
region of the BCS-BEC transition  for the mean-field Leggett model.

\begin{acknowledgments}
The present work has been partially supported by the Hungarian Research
National Foundation under Grant Nos. OTKA T046129 and T038202.
The authors would like to thank useful discussions with 
Prof. P. Sz\'epfalusy and J. Cserti.
\end{acknowledgments}


\begin{thebibliography}{xx}
\bibitem{Jin97} D. S. Jin, J. R. Ensher, M. R. Matthews, C. E. Wieman, and
                E. A. Cornell, Phys. Rev. Lett. {\bf 77}, 420 (1996); D. S.
                Jin, M. R. Matthews, J. R. Ensher, C. E. Wieman, and
                E. A. Cornell, {\it ibid}. {\bf 78}, 764 1997).
\bibitem{Mewes96} D. M. Stamper-Kurn, H.-J. Miesner, S. Inouye, M. R. Andrews,
                and W. Ketterle, Phys. Rev. Lett. {\bf 81}, 500 1998).
\bibitem{kinast04} J. Kinast, A. Turlapov, and J. E. Thomas,
                  Phys. Rev. A {\bf 70}, 051401(R) (2004).
\bibitem{bartenstein042}  M. Bartenstein, A. Altmeyer, S. Riedl,
                  S. Jochim, {\it et al.,}, Phys. Rev. Lett.
                  {\bf  92}, 203201 (2004).
\bibitem{Stringari96} S. Stringari, 
                  Phys. Rev. Lett. {\bf 77}, 2360 (1996).
\bibitem{Stringari04} S. Stringari, 
                  Europhys. Lett. {\bf 65}, 749 (2004).
\bibitem{leggett80} A. J. Leggett, Journal de Physique, Colloque C7,
                  {\bf 41}, 19 (1980), A. J. Leggett,
                  {\it Modern Trends in the Theory of Condensed Matter},
                  (Springer-Verlag, Berlin, 1980),   pp. 13.
\bibitem{Engelbrecht97} J. R. Engelbrecht, M. Randeria, and
                  C. A. R. S\'a de Melo, Phys. Rev. B {\bf 55}, 15153 (1997).
\bibitem{Marini98} M. Marini, F. Pistolesi and G. C. Strinati,
                  Eur.\ Phys.\ J. {\bf 1}, 151 (1998).
\bibitem{Fuchs03} J. N. Fuchs, X. Leyronas, and R. Combescot,
              Phys. Rev. A {\bf 68}, 043610 (2003).
\bibitem{Heiselberg04} H. Heiselberg, 
                  Phys. Rev. Lett. {\bf 93}, 040402 (2004).
\bibitem{Hu04} Hui Hu, A. Minguzzi, Xia-Ji Liu, and M. P. Tosi
              Phys. Rev. Lett. {\bf 93}, 190403 (2004).
\bibitem{Bulgac05} A. Bulgac and G. F. Bertsch, 
                  Phys. Rev. Lett. {\bf 94}, 070401 (2005).
\bibitem{Kim04} Y. E. Kim and A. L. Zubarev,
                  Phys. Rev. A {\bf 70}, 033612 (2004).
\bibitem{Astrakharchik05}
                  G.E. Astrakharchik, R. Combescot, X. Leyronas and
                  S. Stringari,
                  Phys. Rev. Lett.\ {\bf 95}, 030404 (2005).
\bibitem{hermit} The matrix element can be written as
$\langle \Psi_1 |\hat{G}_\Psi |\Psi_2 \rangle= 
  \int d^3r \, (n_0/m)(\nabla \Psi_1^*\sqrt{A_0/n_0})\cdot
  (\nabla \Psi_2 \sqrt{A_0/n_0})$, provided partial integration is 
allowed without a surface term. Usually this is ensured because $n_0$ 
vanishes at the boundary.
\bibitem{Csordas99} A. Csord\'as, and R. Graham,
                  Phys. Rev. A {\bf 59}, 1477 (1999).
\bibitem{Csordas01} A. Csord\'as, and R. Graham,
                  Phys. Rev. A {\bf 64}, 013619 (2001).
\bibitem{Adamunp} Z. Adam, A. Csord\'as, unpublished.
\bibitem{Csordas00} A. Csord\'as, and R. Graham,
                  Phys. Rev. A {\bf 63}, 013606 (2000).
\bibitem{Bruun99} G. M. Bruun, C. W. Clark,
              Phys. Rev. Lett. {\bf 83}, 5415 (1999).
\end{thebibliography}
\end{document}